\documentclass[a4paper,12pt]{article}

\usepackage{fullpage}
\usepackage[T1]{fontenc}

\usepackage{amssymb,amsmath}

\usepackage{graphicx}

\usepackage{amsmath,amsfonts,epsfig,mathrsfs,todonotes,yfonts,cite}  
\newcommand{\nn}{\nonumber}

\newcommand{\re}[1] {(\ref{#1})}

\newcommand{\al}{\alpha} 
\newcommand{\be}{\beta}

\newcommand{\de}{\delta}

\newcommand{\beq}{\begin{equation}}
\newcommand{\eeq}{\end{equation}}

\newcommand{\ber}{\begin{eqnarray}}
\newcommand{\eer}[1]{\label{#1}\end{eqnarray}}
\newcommand{\eero}{\end{eqnarray}}

\newcommand{\half}{{\textstyle{\frac12}}}

\newcommand{\pa}[1]{\partial_{#1}}

\begin{document}

\setcounter{footnote}{0}

\baselineskip 6 mm

\begin{titlepage}
	\thispagestyle{empty}
	\begin{flushright}
		
	\end{flushright}
\rightline{{UUITP-08/26}}

	\vspace{35pt}
	
	\begin{center}
	    {{\Large\bf Symmetries of tensionless strings} }

		\vspace{50pt}
		
		{Ulf Lindstr\"om$^{a,b,c}$}   
		
		\vspace{25pt}

		\vspace{15pt}

        $^{(a)}${\it Department of Physics and Astronomy, Division of Theoretical Physics, Uppsala University, \\ 
        Box 516, SE-75120 Uppsala, Sweden}
		
		\vspace{15pt}

        $^{(b)}${\it Center for Geometry and Physics, Uppsala University, Box 480, SE-75106 Uppsala, Sweden}
		
		\vspace{15pt}      
		
		$^{(c)}${\it Physics Division, National Technical University of Athens \\
        15780 Zografou Campus, Athens, Greece}
  \end{center}
		
		\vspace{40pt}
		
		{ABSTRACT} \\
		
\noindent	In a recent article, \cite{Sheikh-Jabbari:2026cnj} it is stated that a certain scale transformation has been ``systematically overlooked'' in discussions of the tensionless string. Here we point out that this kind of symmetry is treated in numerous places, in the  classical as well as the quantum theory. 

\vspace{\fill}

	\noindent{\footnotesize{Email: ulf.lindstrom@physics.uu.se
 }}

\vspace{10pt}

\bigskip

\end{titlepage}
In an interesting  recent paper \cite{Sheikh-Jabbari:2026cnj} a local scale transformation of the spacetime coordinates compensated by a transformation of he world sheet vector densities is discussed. This is not new, it was introduced in \cite{Isberg:1993av} as part of the extension of the Poincar\' e to conformal symmetry. In  \cite{Sheikh-Jabbari:2026cnj} it is extended to a local symmetry. This is not new either, it is introduced in the Hamiltonian BRST treatment of the conformal string \cite{Gustafsson:1994kr}. However, the discussion in \cite{Sheikh-Jabbari:2026cnj} introduces some interesting aspects, in particular with regard to the constraint structure.  As a complement to that article and a rebuttal of the statement that this kind of symmetry has been  {\em systematically omitted from all prior analyses of the null string}, in this brief  note we emphasise some of the previous work

The tensionless (null) string has been studied in a large number of papers ever since its conception in \cite{Schild:1976vq} and \cite{Karlhede:1986wb}.  Both its bosonic \cite{Lindstrom:1993yb,Gustafsson:1994kr, Isberg:1993av, Isberg:1992ia}
versions and its super versions
\cite{Lindstrom:1990qb, Lindstrom:1990ar}. It was noted very early that the string {\em Weyl invariance gets replaced by conformal invariance } in the tensionless limit\footnote{A quote from many of our articles.} . The purportedly ``overlooked'' symmetry addressed in \cite{Sheikh-Jabbari:2026cnj} is a local version of the  scale transformation in this conformal group. Here we illustrate how this kind of transformations has played important roles in the analysis of the quantum theory in two cases.\\

In more detail, the action for the bosonic tensionless string is \cite{Isberg:1993av}
\beq\label{act}
S=\int d^2\xi V^\al V^\be\pa{\al} X^m\pa{\be} X^n\eta_{mn}
\eeq
where $X^m$ are coordinates of the ambient space time,  $\xi^\al=(\tau.\sigma)$  are coordinates on the world  sheet, and $V^\al$ are two $2d$ vector densities. This model has  $2d$ diffeomorphism invariance along with  conformal symmetries extending the Poincar\'e algebra by \cite{Isberg:1993av}:
\begin{align}\label{confal}\nn
&\de_b X^m=[b\cdot k,X^m]=(b\cdot X)X^m-\half X^2 b^m\\\nn
&\de_b V^\al=-b\cdot X V^\al\\\nn
&\de_aX^m=[as,X^m]=aX^m\\
&\de_a V^\al=-aV^\al~,
\end{align}
where $k^m$ is the generator of conformal boosts, $s$ is the generator of dilatations and $a$ and $b$ are parameters. All of these are  symmetries of the action \re{act}, in particular the last two, which are the transformations considered in \cite{Sheikh-Jabbari:2026cnj} for the case when $a=a(\sigma)$. The conformal algebra \re{confal} (with $a$ constant)  plays a crucial role in the quantum (BRST) analysis of the spectrum of the quantised tensionless string which was found to be topological for spacetime dimensions $d>2$ in \cite{Isberg:1993av}, see also \cite{Isberg:1992ia}.

The quantum analysis was performed in a light cone gauge and misses the structure in  $d=2$ spacetime dimensions. This is amended in \cite{Gustafsson:1994kr} where the conformal string is considered. This is named in analogy to the conformal particle \cite{Marnelius:1978fs} and is described via an embedding in a $d+2$ dimensional space of signature $(-+++...+-)$. The action is
\beq\label{Hamact}
S=\int d^2\xi \big(V^\al V^\be {\cal D}_{\al} X^M{\cal D}_{\be} X^N\eta_{MN}+\Phi X^2\big)
\eeq
where the metric is
\beq
\left( \begin{array}{lll}
\eta_{mn}&0&0\\
0&1&0\\
0&0&-1
\end{array}
\right)
\eeq
and $\Phi$ is a Lagrange multiplier restricting the motion to the higher dimensional light cone (c.f. the construction in \cite{Karlhede:1986wb}). The covariant derivatives are
\ber
{\cal D}_{\al} =\pa{\al}+W_\al
\eer
where $W_\al$ is the gauge field for scale transformations.
Integrating out the $W$ field  and fixing a scaling gauge brings back the action \re{act}.

The action \re{Hamact} has a large number of symmetries: Diffeomorphisms, Scale transformations, Rotations and ``additional'' symmetries. Here we only list the scale transformations
\begin{align}\nn
&\de X^M=\lambda X^M\\\nn
&\de V^\al=-\lambda V^\al\\\nn
&\de W_\al=-\pa{\al}\lambda\\
&\de\Phi=-2\lambda \Phi
\end{align}
where $\lambda$ is a $2d$ scalar field. Again the two first are of the ``overlooked'' type, now with a local parameter. The constraints of this theory are
\beq
\phi^{-1}=P^2~,~~\phi^{0}=P_MX^M~,~~\phi^{1}=X^2~,~~ \phi^L=P_MX'^M
\eeq
where prime denotes derivative with respect to $\sigma$.  Their Poisson brackets form a semi direct product between the Virasoro algebra and a SU (1, 1) Ka\v c-
Moody algebra, both without central extensions, although central extensions are also discussed  in  \cite{Gustafsson:1994kr}. The subalgebra
formed by $\phi^L$ and $\phi^{-1}$ is isomorphic to the gauge algebra in the Minkowski
formulation of the tensionless string. 
The algebra of the Fourier modes of the constraints is also  treated in  \cite{Gustafsson:1994kr}. The Hamiltonian for this theory is the sum of these constraints, each multiplied by its Lagrange multiplier. A BRST quantisation of the model corroborates the results for $d>2$ and yields a critical dimension $d=2$. For details see \cite{Gustafsson:1994kr}.

It should be clear from this brief note that the scale transformations of the form discussed in \cite{Sheikh-Jabbari:2026cnj} have in fact been quite extensively studied, somewhat contradicting the statement in \cite{Sheikh-Jabbari:2026cnj} that this symmetry {\em
 has been systematically omitted from all prior analyses of the null string}. In particular  the conformal string has the full gauged version.
 \bigskip
 
\noindent
{\bf Acknowledgement}:

\noindent
I am grateful to all my collaborators over the years on the some twenty papers on various aspects of tensionless strings. I further thank {\"O}zg{\"u}r Sar{\i}o\u{g}lu for bringing \cite{Sheikh-Jabbari:2026cnj} to my attention.


\begin{thebibliography}{99}


%\cite{Sheikh-Jabbari:2026cnj}
\bibitem{Sheikh-Jabbari:2026cnj}
M.~M.~Sheikh-Jabbari and H.~Yavartanoo,
``On the Consistency of Null Strings Literature: The Tale of an Overlooked Symmetry,''
[arXiv:2605.12414 [hep-th]].
%0 citations counted in INSPIRE as of 24 May 2026

%\cite{Schild:1976vq}
\bibitem{Schild:1976vq}
A.~Schild,
``Classical Null Strings,''
Phys. Rev. D \textbf{16} (1977), 1722
doi:10.1103/PhysRevD.16.1722
%354 citations counted in INSPIRE as of 24 May 2026

%\cite{Karlhede:1986wb}
\bibitem{Karlhede:1986wb}
A.~Karlhede and U.~Lindstr\"om,
``The Classical Bosonic String in the Zero Tension Limit,''
Class. Quant. Grav. \textbf{3} (1986), L73-L75
doi:10.1088/0264-9381/3/4/002
%117 citations counted in INSPIRE as of 24 May 2026

%\cite{Lindstrom:1993yb}
\bibitem{Lindstrom:1993yb}
U.~Lindstr\"om,
``The Zero tension limit of strings and superstrings,''
[arXiv:hep-th/9303173 [hep-th]].
%9 citations counted in INSPIRE as of 24 May 2026

%\cite{Gustafsson:1994kr}
\bibitem{Gustafsson:1994kr}
H.~Gustafsson, U.~Lindstr\"om, P.~Saltsidis, B.~Sundborg and R.~van Unge,
``Hamiltonian BRST quantization of the conformal string,''
Nucl. Phys. B \textbf{440} (1995), 495-520
doi:10.1016/0550-3213(95)00051-S
[arXiv:hep-th/9410143 [hep-th]].
%45 citations counted in INSPIRE as of 24 May 2026

%\cite{Isberg:1993av}
\bibitem{Isberg:1993av}
J.~Isberg, U.~Lindstr\"om, B.~Sundborg and G.~Theodoridis,
``Classical and quantized tensionless strings,''
Nucl. Phys. B \textbf{411} (1994), 122-156
doi:10.1016/0550-3213(94)90056-6
[arXiv:hep-th/9307108 [hep-th]].
%186 citations counted in INSPIRE as of 24 May 2026

%\cite{Isberg:1992ia}
\bibitem{Isberg:1992ia}
J.~Isberg, U.~Lindstr\"om and B.~Sundborg,
``Space-time symmetries of quantized tensionless strings,''
Phys. Lett. B \textbf{293} (1992), 321-326
doi:10.1016/0370-2693(92)90890-G
[arXiv:hep-th/9207005 [hep-th]].
%52 citations counted in INSPIRE as of 25 May 2026


%\cite{Lindstrom:1990ar}
\bibitem{Lindstrom:1990ar}
U.~Lindstr\"om, B.~Sundborg and G.~Theodoridis,
``The Zero tension limit of the spinning string,''
Phys. Lett. B \textbf{258} (1991), 331-334
doi:10.1016/0370-2693(91)91094-C
%58 citations counted in INSPIRE as of 24 May 2026

%\cite{Lindstrom:1990qb}
\bibitem{Lindstrom:1990qb}
U.~Lindstr\"om, B.~Sundborg and G.~Theodoridis,
``The Zero tension limit of the superstring,''
Phys. Lett. B \textbf{253} (1991), 319-323
doi:10.1016/0370-2693(91)91726-C
%75 citations counted in INSPIRE as of 24 May 2026

%\cite{Marnelius:1978fs}
\bibitem{Marnelius:1978fs}
R.~Marnelius,
``Manifestly Conformal Covariant Description of Spinning and Charged Particles,''
Phys. Rev. D \textbf{20} (1979), 2091
doi:10.1103/PhysRevD.20.2091
%74 citations counted in INSPIRE as of 25 May 2026


\end{thebibliography}
\end{document}